\documentclass[twocolumn,showpacs,preprintnumbers,amsmath,amssymb]{revtex4}
\usepackage{graphicx}
\usepackage{dcolumn}
\usepackage{bm}
\begin{document}

\title{Tuning of the Gap in a Laughlin-Bychkov-Rashba Incompressible Liquid}
\author{Marco Califano$^{\ast}$, Tapash Chakraborty$^{\ast\ddag}$ 
and Pekka Pietil\"ainen$^{\ast\dag}$}
\affiliation{$^\ast$Department of Physics and Astronomy,
University of Manitoba, Winnipeg, Canada R3T 2N2}
\affiliation{$^\dag$Department of Physical Sciences/Theoretical Physics,
P.O. Box 3000, FIN-90014 University of Oulu, Finland} 
\date{\today}
\begin{abstract}
We report on our investigation of the influence of Bychkov-Rashba 
spin-orbit interaction (SOI) on the incompressible Laughlin state.
We find that experimentally obtainable values of the spin-orbit 
coupling strength can induce as much as a 25\% increase in the 
quasiparticle-quasihole gap $E_g$ at low magnetic fields in InAs, 
thereby increasing the stability of the liquid state. The 
SOI-modulated enhancement of $E_g$ is also significant for 
$\nu=1/5$ and $1/7$, where the FQH state is usually weak. This 
raises the intriguing possibility of tuning, via the SO coupling 
strength, the liquid to solid transition to much lower densities.
\end{abstract}
\maketitle

The ground state of a two-dimensional electron gas (2DEG) in 
the presence of a high perpendicular magnetic field is well 
known to appear as a multitude of highly correlated incompressible 
fractional quantum Hall effect (FQHE) states at a few  special
values of the Landau level filling factors \cite{laughlin,book}. 
It is also well established that the strongest effect appears 
at the lowest Landau level filling factor ($\nu=\frac13$) that has the 
largest quasiparticle-quasihole gap \cite{laughlin,book}. The gaps 
are much smaller for $\nu < \frac13$, and as a consequence, nature 
of the electron states in that regime has remained a challenge until 
now. This is due to the fact that in the low density regime, the
incompressible liquid is expected to undergo a phase transition to
a crystalline state \cite{lowdense}. However, a definitive conclusion
about the onset of this quantum phase transition has remained elusive
because experimentally one observes weak effects for filling factors
$\frac15, \frac17$, etc., and theoretically, one compares two very
small energies (ground state) in order to determine which phase is 
energetically favored. For more than two decades, investigations of 
the FQHE have focused largely on 2DEGs that are embedded in GaAs 
heterostructures, where spin-related effects are small (though 
important \cite{advances}) compared to other effects, because of the 
small value of the $g$-factor of electrons in GaAs \cite{halperin}. 
However, studies of the spin-orbit (SO) coupling in a 2DEG within 
an InAs (or InSb) quantum well with very large $g$ values, are at 
the cusp of a rapid advance, due largely to their relevance to 
spin transport in low-dimensional electron channels \cite{grundler}. 
In order to investigate the influence of SO interaction on the 
incompressible Laughlin states, we have carried out the well 
established finite-size studies in a periodic rectangular 
geometry, but for the first time with the SO coupling included in
the Hamiltonian. We find that as the SO coupling strength
is increased, there is an increase in the quasiparticle-quasihole 
gap, indicating that the incompressible state can be rendered more 
stable by appropriate tuning of the SO coupling. This might prove 
to be particularly useful in the low-electron density regime where, as
described above, the FQHE state is usually very weak for conventional 
systems.

For a 2DEG in the $xy$ plane with a magnetic field {\bf B} along 
the $z$ direction, in the Landau gauge [with the choice of vector 
potential ${\bf A}=(0,Bx,0)$], the single-particle states and the 
corresponding energies are obtained by solving the one-electron 
Hamiltonian
\begin{equation}
H = \frac{({\bf p}+e{\bf A})^2}{2m^*}+\frac{\alpha}{\hbar}
[\mbox{\boldmath $\sigma$}\times({\bf
p}+e{\bf A})]_z+g\mu_BB\sigma_z
\end{equation}
that includes the Bychkov-Rashba term \cite{rashba}
and the Zeeman term. Here ${\bf p}$ is the momentum operator, 
$\alpha$ is the SO coupling strength and 
$\mbox{\boldmath $\sigma$}=(\sigma_x, \sigma_y, \sigma_z)$
are the Pauli spin matrices. Experimental values of the SO coupling 
constant lie in the range of 5--45 meV$\cdot$nm \cite{expt}. The 
high values of $\alpha$ are deduced from magnetotransport experiments
where the SO interaction is tuned for a fixed carrier density in 
a 2DEG by using gate electrodes in a square asymmetric InAs 
quantum well \cite{expt}. We will therefore focus our investigation 
on InAs, as it represents the most promising 
material, as yet, for achieving large SO coupling.

We solve the Schr\"odinger equation
\begin{equation}
H\psi = E\psi\label{eq:schr}
\end{equation}
in a rectangular geometry with supercell sides $L_x$ and $L_y$ 
(i.e., aspect ratio $\lambda=L_x/L_y$) and expand the single-particle  
wavefunctions $\psi_{k_y}({\bf r})$ as a superposition of solutions 
of the Hamiltonian in the absence of SO interaction \cite{mag_field}
\begin{equation}
\psi_{k_y}({\bf r}) =
e^{ik_yy}\sum_{n,\sigma}\phi_n(x-x_0)C^{\sigma}_n\,|\sigma\rangle/
\sqrt{L_y}\label{eq:psi_sp}
\end{equation}
where 
$$\phi_n(x-x_0)=\beta_n e^{-(x-x_0)^2/2l_0^2}H_n[(x-x_0)/l_0]/
\sqrt{\sqrt{\pi}l_0}$$
is the usual solution to the harmonic oscillator problem, with $H_n(x)$ 
the Hermite polynomial,
$l_0=(\hbar/m^*\omega_c)^{1/2}$ the radius of the cyclotron orbit
with frequency $\omega_c=eB/m^*$ and center $x_0=k_yl_0^2$, 
$n$ is the Landau level index, $\beta_n=1/\sqrt{2^nn!}$,
and
$$\sigma = up,dn = \left(\begin{array}{c}1\\0\end{array}\right),
\left(\begin{array}{c}0\\1\end{array}\right)$$
is the electron spinor.

\begin{figure}[b]
\begin{center}
 \includegraphics[width=.45\textwidth]{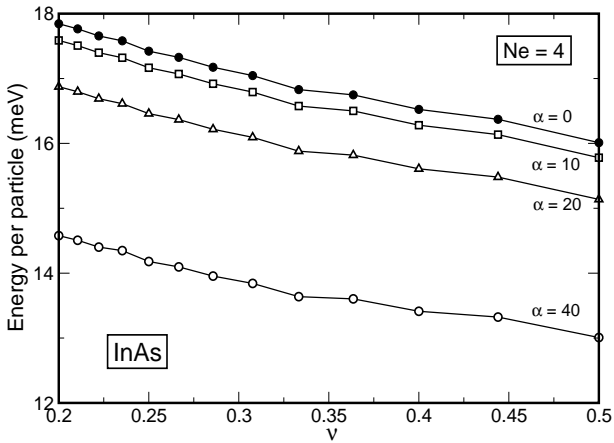}
\protect\caption{
Ground state energy per particle as a function of filling factor 
$\nu$ for 4 different values of the SO coupling constant 
$\alpha=0,10,20,40$ calculated for 4 electrons in the lowest two 
Landau levels in InAs ($B=10$ Tesla). 
}\label{fig:E0_vs_nu}
\end{center}
\end{figure}
Substituting Eq. (\ref{eq:psi_sp}) into the Schr\"odinger 
equation (\ref{eq:schr}), multiplying both
sides by $\phi_l(x-x_0)$ and
integrating over $x$, we obtain a system of equations \cite{mag_field}
\begin{eqnarray}
&i(\alpha/l_0)\sqrt{2l}C_{l-1}^{up}+[(l+1/2)\hbar
\omega_c+E_d]C_l^{dn} = 0&\nonumber\\
&[(l+1/2)\hbar\omega_c+E_u]C_l^{up}-i(\alpha/l_0)
\sqrt{2(l+1)}C_{l+1}^{dn} =  0&\nonumber\\
&l=0,1,2,\ldots&
\end{eqnarray}
whose solution yields \cite{mag_field}
$$(1/2\hbar\omega_c+E_d)C_s^{dn}=0, \,\,\,\,s=0,$$
$$ \left[\begin{array}{cc}
(s-1/2)\hbar\omega_c+E_u & -i(\alpha/l_0)\sqrt{2s} \\
i(\alpha/l_0)\sqrt{2s} & (s+1/2)\hbar\omega_c+E_d
\end{array}\right]
\left(\begin{array}{c}
C_{s-1}^{up} \\ C_{s}^{dn}
\end{array}\right) =0, $$
$s=1,2,3,\ldots$, where $E_{u}=g\mu_BB-E$ and $E_{d}=-g\mu_BB-E$.
Corresponding to $s=0$ there is only one level, the same as the 
lowest Landau level without SO interaction, with energy
$$E_0=1/2\hbar\omega_c-g\mu_BB$$
and wavefunction
$$\psi_{0,k_y}=e^{ik_yy}\phi_0(x-x_0)/\sqrt{L_y}
\left[\left(\begin{array}{c}
1 \\ 0
\end{array}\right)+
\left(\begin{array}{c}
0 \\ 1
\end{array}\right)
\right].$$
For all other values of $s\neq 0$ there are two branches of levels
\cite{mag_field}
\begin{equation}
\psi_{s,k_y}^+=\frac{e^{ik_yy}}{\sqrt{L_yA_s}}
\left(\begin{array}{c}
-iD_s\phi_{s-1}(x-x_0)\\ \phi_s(x-x_0)
\end{array}\right)\label{eq:psi_p}
\end{equation}
\noindent and
\begin{equation}
\psi_{s,k_y}^-=\frac{e^{ik_yy}}{\sqrt{L_yA_s}}
\left(\begin{array}{c}
\phi_{s-1}(x-x_0)\\ -iD_s\phi_s(x-x_0)
\end{array}\right)\label{eq:psi_m}
\end{equation}
with energies
\begin{equation}
E_s^{\pm}=s\hbar\omega_c\pm\sqrt{E_0^2+2s\alpha^2/l_0^2}.\label{eq:Ekin} 
\end{equation}
Here
\begin{equation}
D_s=\frac{\sqrt{2s}\alpha/l_0}{E_0+\sqrt{E_0^2+2s\alpha^2/l_0^2}}\label{eq:D}
\end{equation}
and $A_s=1+D_s^2$. From Eq. (\ref{eq:psi_p}) and (\ref{eq:psi_m}) 
we see that SO interaction couples {\em two} Landau levels. 
While previous works \cite{mag_field} were restricted to the study 
of the single-particle states, we use these equations as a starting 
point for our {\em exact} many-body treatment.
Applying periodic boundary conditions, we obtain ($k_y=x_0/l_0^2$)
$$ x_0=X_j=2\pi l_0^2j/L_y, \ L_x=2\pi l_0^2m/L_y, $$
and consequently,
\begin{eqnarray*}
\psi_{s,j}^+({\bf r}) &=&
\frac1{\sqrt{\sqrt{\pi}l_0L_yA_s}}\sum_{n}\exp\left[i(X_j+nL_x)
\frac{y}{l_0^2}\right. \\
&-&\left.\frac{(X_j+nL_x-x)^2}{2l_0^2}\right] \\
&&\times\left(\begin{array}{c}
-iD_s\beta_{s-1}H_{s-1}\left(\frac{(X_j+nL_x-x)}{l_0}\right)\\
\beta_{s}H_{s}\left(\frac{(X_j+nL_x-x)}{l_0}\right)
\end{array}\right) \\
\psi_{s,j}^-({\bf r}) &=&
\frac1{\sqrt{\sqrt{\pi}l_0L_yA_s}}\sum_{n}\exp\left[i(X_j+nL_x)
\frac{y}{l_0^2}\right. \\
&-&\left.\frac{(X_j+nL_x-x)^2}{2l_0^2}\right] \\
&&\times \left(\begin{array}{c}
\beta_{s-1}H_{s-1}\left(\frac{(X_j+nL_x-x)}{l_0}\right)\\
-iD_s\beta_{s}H_{s}\left(\frac{(X_j+nL_x-x)}{l_0}\right)
\end{array}\right).
\end{eqnarray*}
We then build the antisymmetrized products (Slater determinants) 
using $\psi^+$ and $\psi^-$ as a complete basis
for the many-body wavefunction expansion
$$\Psi=\sum_{\{i_k\}}{\mathcal{P}}(i_1,i_2,\ldots,i_n)
a^{\dagger}_{i_1}a^{\dagger}_{i_2}\ldots
a^{\dagger}_{i_n}|0\rangle$$
where $i_k=(s_k,j_k,\tilde{\sigma}_k)$, $\tilde{\sigma}_k=\pm$ and 
${\mathcal{P}}(i_1,i_2,\ldots,i_n)$ is the
antisymmetrization operator.

The many-body Schr\"odinger equation was then solved by performing 
an exact diagonalization of the many-body Hamiltonian
\begin{equation}
{\mathcal{H}}=\sum_j{\mathcal{W}}_j\,a^{\dagger}_ja_j+\sum_{j_1}
\sum_{j_2}\sum_{j_3}\sum_{j_4}{\mathcal{A}}_{j_1j_2j_3j_4}
a^{\dagger}_{j_1}a^{\dagger}_{j_2}a_{j_3}a_{j_4}\label{eq:Schr_mb}.
\end{equation}
The kinetic energy term 
\begin{equation}
{\mathcal{W}}_j={\mathcal{S}}+E_j\label{eq:onebody}
\end{equation}
includes the effects of a neutralizing background. 
Here $E_j=E_s^{\pm}$,
\begin{equation}
{\mathcal{S}}=-\frac{e^2}{\epsilon l_0}\frac{1}{\sqrt{2\pi
m}}\left[2-\sum_{k_1,k_2}'\sqrt{\frac{\pi}{z}}(1-\mbox{erf}
(\sqrt{z}))\right]\label{eq:Emadelung}
\end{equation}
is the Madelung energy, $e$ the electron charge, $\epsilon$ the
dielectric constant, $z=\pi(\lambda^2 k_1^2+k_2^2)/\lambda$, and the
prime in the summation means that the term $k_1=k_2=0$ is excluded.
The expression for the scattering matrix element ${\mathcal{A}}_{i_1i_2i_3i_4}$ 
depends on the quantum numbers $i_1i_2i_3i_4$, where, again, 
$i_k=(s_k,j_k,\tilde{\sigma}_k)$. For the case of positively polarized 
``spins'' (i.e., $\tilde{\sigma}_k=+$ for all $k$) we have:
\begin{widetext}
\begin{eqnarray}
{\mathcal{A}}_{i_1i_2i_3i_4}|_{\tilde{\sigma}_i=+}&=&
\delta'_{j_1+j_2,j_3+j_4}\frac12\frac{e^2}{\epsilon l_0}
\sqrt{\frac{\lambda}{2\pi m}}\prod_{i=1}^4\left(\frac{D_{s_i}
\beta_{s_i}}{A_{s_i}}\right)(-1)^{s_2+s_4}s_1!s_2!s_3!s_4! \nonumber\\
&&\times\sum_{k_1=1}^{\infty}\sum_{k_2=-\infty}^{\infty}
\frac{\delta'_{j_1-j_4,k_2}}{\sqrt{k_1^2+\lambda^2k_2^2}}
e^{-\pi(k_1^2+\lambda^2k_2^2)/\lambda m}\left(k_2\sqrt{\frac{2\pi
\lambda}{m}}\right)^{\sum_{i=1}^4s_i}\cos{\left[
\frac{2\pi}{m}k_1(j_1-j_3)\right]}\nonumber\\
&&\times\left\{\sum_{t_1=0}^{s_1}\sum_{p_1=1}^{up_1^c}
\sum_{t_2=0}^{s_3}\sum_{p_2=1}^{up_2^c}
\frac{s_1!}{(s_1-t_1)!}\frac{s_3!}{(s_3-t_2)!}
\left(
\begin{array}{c}
s_4 \\ t_1+2p_1
\end{array}\right)
\left(
\begin{array}{c}
s_2 \\ t_2+2p_2
\end{array}\right)
(-1)^{t_1+t_2+p_1+p_2}\right.\nonumber\\
&&\times\left.2^{t_1+t_2-1}\left(k_2\sqrt{
\frac{2\pi\lambda}{m}}\right)^{2(t_1+p_1+t_2+p_2)}
\left(k_1\sqrt{\frac{2\pi}{\lambda m}}\right)^{2(p_1+p_2)}{
\mathcal{L}}_{t_1}^{2p_1}\left(k_1^2\frac{2\pi}{\lambda 
m}\right){\mathcal{L}}_{t_2}^{2p_2}\left(k_1^2\frac{2\pi}{\lambda
m}\right)
\right\}\label{eq:aijkl}
\end{eqnarray}
\end{widetext}
\noindent where
\begin{equation*}
{\mathcal{L}}_{n}^{\alpha}(x)=\sum_{m=0}^n (-1)^m
\left(
\begin{array}{c}
n+\alpha \\ n-m
\end{array}\right)
\frac{x^m}{m!}
\end{equation*}
are the Laguerre polynomials and 
\begin{equation*}
up_1^c= \left\{ \begin{array}{ll}
                  (s_4-t_1) & \mbox{if $(s_4-t_1)<0$} \\
                  \mbox{Int}\{(s_4-t_1)/2\} & \mbox{otherwise} \\
                \end{array}
        \right.
\end{equation*}
\noindent
and the same yields for $up_2^c$, with $s_4\rightarrow s_2$ and
$t_1\rightarrow t_2$, [Int$\{x\}$ is the integer part of $x$].

Due to the presence of the spinors $\sigma$, the two branches 
$\psi^{\pm}$ and the coupling of the Landau levels in 
pairs, the derivation of the complete expression for 
${\mathcal{A}}_{i_1i_2i_3i_4}$ is highly nontrivial. Moreover, 
the Hamiltonian matrix can be very large and its diagonalization 
computationally very intensive. We calculated the ground 
state energy per particle $E_0$ for a system with four 
electrons in the lowest two Landau levels with different filling 
factors $\nu=4/N_s$, where $N_s=8,9,\ldots,20$. The results 
are shown in Fig.~\ref{fig:E0_vs_nu} for $B=10$ Tesla. 
We see that the presence of SO coupling lowers considerably 
the value of $E_0$, compared to the result for $\alpha=0$, which 
coincides with the usual results 
obtained for the FQHE with no SO, with both the Zeeman and the 
kinetic energies included. 

\begin{figure}[b]
\begin{center}
 \includegraphics[width=.45\textwidth]{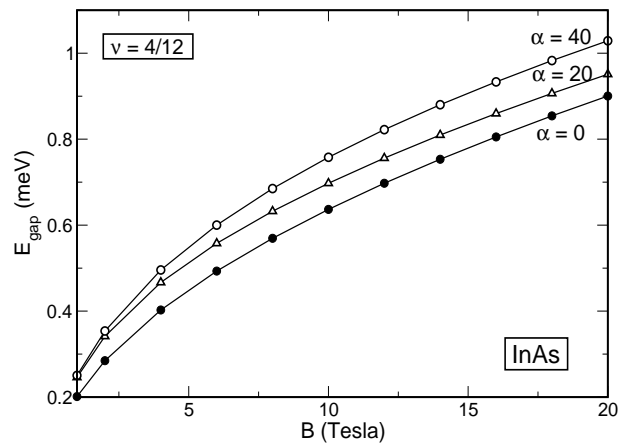}
\protect\caption{
Quasiparticle-quasihole energy gap as a function of magnetic field 
$B$ for a filling factor $\nu=1/3$ and three different values of the 
SO coupling constant $\alpha=0,20,40$ calculated 
for four electrons in the lowest two Landau levels in InAs. 
}\label{fig:Egap412}
\end{center}
\end{figure}

The most intriguing effect of SO coupling in 2DEG, however, 
is found in connection with the magnitude of the 
quasiparticle-quasihole energy gap $E_{g}$, derived from the 
positive discontinuity of the chemical potential at the 
filling factor $\nu$ \cite{book}.

\begin{figure}[h]
\begin{center}
 \includegraphics[width=.45\textwidth]{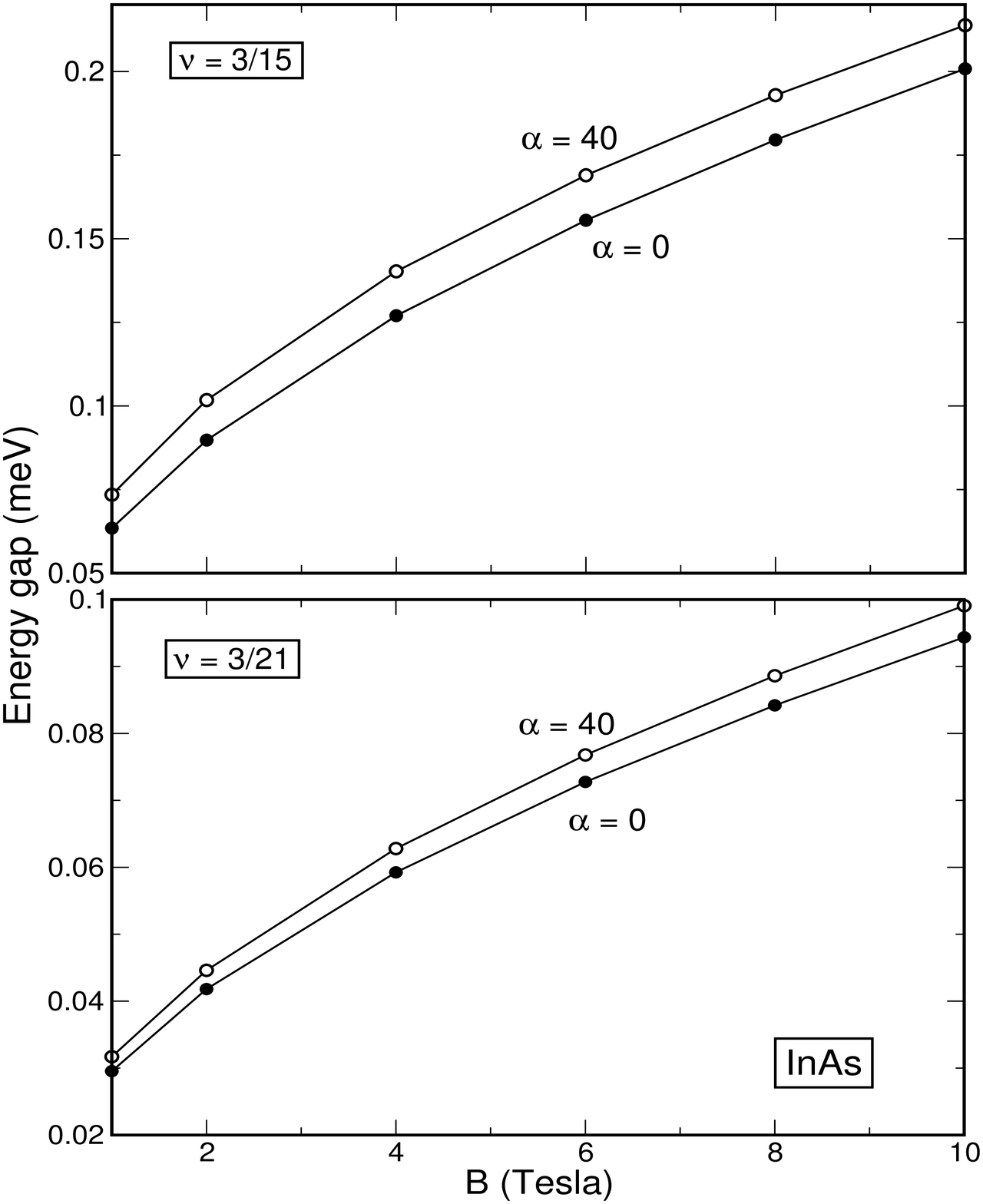}
\protect\caption{
Quasiparticle-quasihole energy gap as a function of magnetic field 
$B$ for a filling factor $\nu=1/5$ and 1/7  
and two different values of the SO coupling constant $\alpha=0,40$ 
calculated for three electrons in the lowest two Landau levels in InAs. 
}\label{fig:Egap3}
\end{center}
\end{figure}
As shown in Fig.~\ref{fig:Egap412} for four electrons and a 
filling factor of $\frac13$ and in Fig.~\ref{fig:Egap3} for 
even smaller values of $\nu=\frac15$ and $\frac17$ 
\cite{note}, large values of $\alpha$ cause the enhancement of 
$E_{g}$. This enhancement is larger for small magnetic fields 
($B\sim 1$ Tesla, i.e., fields for which the Rashba term 
is still comparable to the Zeeman term), and can be of the order of 
25\% for $\nu=\frac4{12}$ and mid-range values of the coupling 
constant ($\alpha=20$). This is seen in Fig.~\ref{fig:Egap412}, 
where $E_{g}$ increases from 0.20 meV for $\alpha=0$ to $\sim 0.25$ 
for $\alpha=20$. Smaller increases occur at smaller filling 
factors, where $E_{g}$ shows a 17\% and 7\% increase for 
$\nu=\frac{3}{15}$ and $\frac{3}{21}$, respectively. 
Reasons for this behaviour derive from a complex interplay 
between the different one- [Eq. (\ref{eq:onebody})] and 
two-body [Eq. (\ref{eq:aijkl})] terms.
For $\alpha\ne0$, one clear effect of the SO interaction is that the
kinetic energy is no longer constant for a given value of the
magnetic field [Eq.~(7)] but depends on the strength of the SO coupling. 
This, coupled with the fact that the interaction terms are profoundly 
modified by the SO interaction, results in a change in 
$\partial E_0/\partial\nu$. This is reflected in an increase of 
the energy gap.

In summary, we have investigated the influence of the SO coupling
(Bychkov-Rashba) on the incompressible state proposed by Laughlin
at $\nu=\frac13$, using the exact diagonalization scheme for 
finite-size systems in a periodic rectangular geometry. We found 
that, as the SO coupling strength is increased, there is an
increase of the quasiparticle-quasihole gap. This is particularly 
advantageous for filling factors $\nu < 1/5, 1/7$ where a larger gap 
would signify a more stable liquid state that will push the 
liquid-solid transition further down in the density. 

The work of T.C. has been supported by the Canada 
Research Chair Program and the Canadian Foundation for Innovation 
Grant. The work of M. C. has been supported by NSERC.


\begin{thebibliography}{99}
\bibitem{note} We used three electrons for the most computationally 
intensive calculations that involved a high number of basis states 
and therefore large matrices, and would otherwise 
have been impossible to perform with four electrons.
\end{thebibliography}
\end{document}